\documentclass[12pt]{iopart}
\usepackage{amssymb}

\newcommand{\Neff}{N_\mathit{eff}}
\newcommand{\cL}{{\mathcal L}}
\newcommand{\N}{{\mathcal N}}
\newcommand{\Dv}{\Delta{\mathit v}_p}
\newcommand{\mF}{\mathbb{F}}

\begin{document}

\title[Structure of density matrix with minimal
uncertainty]{Structure of the density matrix providing the minimum
of generalized uncertainty relation for mixed states}

\jl{1}

\author{M. U. Karelin\footnote[2]{E-mail: karelin@ifanbel.bas-net.by},
\fbox{A. M. Lazaruk}}
\address{Institute of Physics, National\\
Academy of Sciences,\\ Minsk 220072, Belarus}
\begin{abstract}
For configurational space of arbitrary dimension a strict form of
the uncertainty principle has been obtained, which takes into
account the dependence of inequality limit on the effective number
of pure states present in given statistical mixture. It is shown
that in a state with minimal uncertainty the density operator's
eigenfunctions coincide with the stationary wavefunctions of a
multidimensional harmonic oscillator. The mixed state obtained has
a permutational symmetry which is typical for a system of
identical bosons.
\end{abstract}

 \submitted

\section{Introduction}

As shown in the papers [1-3], the generalization of conventional
uncertainty relation of the position-momentum type (see, e. g.
Ref. 4) to the case of mixed states, determined by a density
matrix $\rho(x,x')$, leads to a radically new result. It sounds as
follows --- the smallest allowable room $\Delta {\mathit v}_p$,
occupied by system in the phase space, grows proportionally to the
effective number $N_\mathit{eff}$ of pure states by whose
statistical mixture the given density matrix is representable,
that is
 $$
 \min(\Dv) \propto \Neff.
 \eqno(1)
 $$
One of the possible variants of relation (1), which can be used in
a real or configurational space ${\mathbf X}=(x_1,\dots,x_s)$ with
an arbitrary number of dimensions $s$, has been obtained in [5]
and is of the form
 $$
 \frac{(\Delta x\;\Delta q)^s }{\Neff} \ge
 C(s)\left( \frac{1}{2} \right)^s.
 \eqno(2)
 $$
The next definitions are introduced here: $\Delta x$, $\Delta q$ –
the root mean square widths (additionally averaged over all $s$
degrees of freedom) correspondingly of coordinate and wave-vector
distributions in the state $\rho(x,x')$
 $$
 (\Delta x)^2  = \frac{1}{s}\int{d^s\!X\:\mathbf{X}^2
 \rho (\mathbf{X},\mathbf{X})},
 \eqno(3)
 $$
 $$
 (\Delta q)^2  =  - \frac{1}{s}
 \int {d^s\!X\int {d^s\!X' \,\delta ({\mathbf X} - {\mathbf X'})
 \, \nabla_{\mathbf X}^2 \rho ({\mathbf X},{\mathbf X'})} }
 ,\quad
 \left[ \nabla_{\mathbf X}^2  = \sum\limits_{i = 1}^s
 {\partial_{x_i}^2}\right],
 \eqno(4)
 $$
 $$
 \Neff  = \left(\int d^s\!X \int d^s\!X'\left|
 \rho({\mathbf X},{\mathbf X'})\right|^2\right)^{-1}.
 \eqno(5)
 $$
Besides, it is assumed that the normalisation condition
 $$
 \int d^s\! X\, \rho({\mathbf X},{\mathbf X}) = 1,
 \eqno(6)
 $$
is met and the mean values of position and momentum are equal to
zero (without loss of generality the latter can be done by
choosing an appropriate reference frame). For convenience the
right-hand side of inequality (2) is represented as two factors.
One of which, $(1/2)^s$, is the minimum of usual uncertainty
relation for pure states and, as known, is attained for Gaussian
wave-packets [4]. The second factor, $C(s)$, can be named the
``packing coefficient'' and expresses the specificity of $\Neff$
definition and, properly, the influence of space
multidimensionality [5]
 $$
 C(s) = \frac{2^{s + 1}(s + 1)!}{(s + 2)^{s + 1}}.
 \eqno(7)
 $$
The particular interest for the relation (2) is caused by its
analogy with one of the basic statements of statistical mechanics
[6, 7] concerning the partitioning of the phase space into cells
with each of them corresponds to one quantum state. Indeed, the
quantity $\Dv=(\Delta x,\Delta q)^s$ is a peculiar kind of measure
of a system phase volume, whereas $\Neff$ is a characteristic
number of its possible states. It is therefore obvious that,
despite the considerable differences in the formulations used, in
either case one can speak about two possible ways of exhibiting
one and the same fundamental property of quantum-mechanical
objects.

On the other hand, inequality (2) holds not only for density
matrix, but as well for correlation functions of wavefields of
various nature and it can be seminally applied, after appropriate
terminology corrections, in other fields of physics. In fact, the
basic works [1-3], where one-dimensional case of relation (2) was
obtained, deal with partially coherent light beams.

Given in [5] variant of the proof was based on the Carlson type
integral inequalities for the Wigner function, that implies some
disadvantages, in particular, because such approach does not
determine the explicit form of the density matrix minimizing (2).
This question is of interest in itself and, at the same time, is
important for applications (for example, at the coherence theory
[8]). Hence in the present paper we shall consider an alternative
way to derive the relation (2). The aim is not only to solve the
problem stated, but also reinforces the inequality by taking into
account the dependence of packing coefficient on the effective
number of states $C(s)\Rightarrow C(s,\Neff)$.

\section{Formulation of the problem and rigorous solution}

The method to be applied constitutes a modification of approach,
which was developed in Ref. 9 to analyze an analogous
one-dimensional problem. It is based on the use of standard
Lagrange procedure of search for the minimum of the system
uncertainty volume (i. e. the value $\Delta x\Delta q$) under a
constraint of given $\Neff$ (5). As common for such treatments
[1-3, 5, 9], one should seek the extremum of auxiliary functional
$\mF$
 $$
 \mF[\rho ({\bf X},{\bf X'})] =
 k^2 (\Delta x)^2 + \frac{1}{k^2}(\Delta q)^2,
 \eqno(8)
 $$
which has the property that its minimum for $k$ ($k$ is a variate
scale factor) is attained simultaneously with the minimum of
uncertainty volume
 $$
 \mathop {\min }\limits_k \mF = 2\Delta x\Delta q.
 $$
The substitution of definitions (3), (4) into (8) shows that the
value of $\mF$ coincides in form with the mean value of energy of
$s$-dimensional symmetrical harmonic oscillator with a Hamiltonian
 $$
 \frac{1}{s}\left(-\frac{1}{k^2}\nabla_{\bf X}^2+k^2{\bf X}^2\right),
 $$
which is in the mixed state $\rho({\bf X},{\bf X'})$. It is therefore
natural for further treatment to represent $\rho({\bf X},{\bf X'})$ as
a series expansion in the basis of this Hamiltonian eigenfunctions
 $$
 \rho({\mathbf X},{\mathbf X'})=\sum\limits_{{\mathbf n},{\mathbf n'}}
 {a_{{\mathbf n},{\mathbf n'}}\Psi _{\mathbf n}({\mathbf X})
 \Psi_{\mathbf n'}({\mathbf X'})},
 \eqno(9)
 $$
where ${\mathbf n}=(n_1,...,n_s)$ is the ``vector'' index whose
components range over nonnegative integer numbers,
 $$
 \Psi _{\mathbf n}({\mathbf X})=\prod\limits_{i = 1}^s{\psi_{n_i}(x_i )},
 $$
 $$
 \psi_n(x)=\sqrt{\frac{k}{2^n n!\sqrt\pi}}\exp\left(-k^2 x^2/2\right)
 H_n(kx),
 $$
$\displaystyle
H_n(x)=\frac{1}{\sqrt[4]{\pi}}e^{x^2}\frac{d^n}{dx^n}e^{-x^2}$
--- Hermite polynomials, the functions $\Psi _{\mathbf n}({\mathbf
X})$ are real and orthonormal. In view of the foregoing, the
substitution of (9) into (8) gives
 $$
 \mF = \sum\limits_{\mathbf n}{a_{{\mathbf n},{\mathbf n}}
 \left(\frac{2}{s}\left\|{\mathbf n}\right\|+ 1\right)},\quad
 \left\|{\mathbf n}\right\|=\sum\limits_{i = 1}^s{n_i}.
 \eqno(10)
 $$
Now the additional condition, under which the effective number of
states (5) is constant, can be written in the form
 $$
 \mu  = \frac{1}{N_\mathit{eff}} =
 \sum\limits_{{\mathbf n},{\mathbf n'}}
 {\left|a_{{\mathbf n},{\mathbf n'}}\right|^{2} }={\rm const}
 \eqno(11)
 $$
(in statistical optics the parameter $\mu$ is frequently referred
to as global degree of coherence). One more constraint on
expansion coefficients $a_{{\mathbf n},{\mathbf n'}}$ follows from
the requirement for the density matrix (6) normalization
 $$
 \sum\limits_{\mathbf n}{a_{{\mathbf n},{\mathbf n}}}=1.
 \eqno(12)
 $$

Without going into details of elementary, although cumbersome
intermediate calculations, we now turn to the analysis of the
solution obtained. The first and most important consequence of
minimization of $\mF$ (10) in coefficients $a_{{\mathbf
n},{\mathbf n}}$ is that in the state of least uncertainty
$\rho_\mathit{min}({\mathbf X}, {\mathbf X'})$ all off-diagonal
elements of the matrix $[a_{{\mathbf n},{\mathbf n}}]$ are equal
to zero
 $$
 a_{{\mathbf n},{\mathbf n'}}=a_{{\mathbf n},{\mathbf n}}
 \delta_{{\mathbf n},{\mathbf n'}}.
 $$
It means, that in this case the density operator's eigenfunctions
[6] (in optics --- the decomposition modes) coincide in form with
the eigenfunctions of energy operator of multidimensional harmonic
oscillator. Accordingly, the diagonal elements of matrix
$[a_{{\mathbf n},{\mathbf n'}}]$ are the eigenvalues of density
operator
 $$
 a_{{\mathbf n},{\mathbf n}}\Psi_{\mathbf n}({\mathbf X})=
 \int{d^s X'\rho_\mathit{min}({\mathbf X},{\mathbf X'})
 \Psi_{\mathbf n}({\mathbf X'})}.
 $$
They define the probabilities to find the system in the pure state
$\Psi_{\mathbf n}({\mathbf X})$ and satisfy the conditions
${\mathop{\rm Im}\nolimits}\: a_{{\mathbf n},{\mathbf n}} = 0$ (as
a consequence of $[a_{{\mathbf n},{\mathbf n'}}]$ is Hermitian)
and $0\le a_{{\mathbf n},{\mathbf n}} \le 1$.

It should be noted here that the use of characteristic $\Neff$ as
a measure of number of possible pure states of the system is
directly related to the operation of transition to the basis of
density operator's eigenfunctions in the representation (9). In
the theory of stochastic processes [10] this procedure is referred
to as Karhunen-Lo\'eve expansion of ``correlation function''
$\rho({\mathbf X}, {\mathbf X'})$ and gives, as known, the most
compact and the most rapidly converging form of this series.
Therefore, in this paper (as well as in [5]) under the term
``state'' is meant, as a rule, the eigenstate of density operator.

The second conclusion, directly following from the form of (10) --
(12), is that the weights of states $a_{{\mathbf n},{\mathbf n}}$
can only depend on the vector index norm $\|{\mathbf n}\|$, and
this dependence is linear and decreasing when $\|{\mathbf n}\|$
increases. Then it is obvious that at any finite value of $\Neff$
the number of pure states $\N$ present in the expansion
$\rho_\mathit{min}$ with a probability other then zero is also
finite. And in the index domain, the coefficients $a_{{\mathbf
n},{\mathbf n}} \ne 0$ fill, layer by layer, the interior of $s$%
-dimensional equilateral pyramid with the total number of layers
equal to $\cL$ ($0 \le \|{\mathbf n}\| < \cL$). The most probable
pure state (${\mathbf n}=0$) corresponds to the pyramid's vertex.
The terms of expansion (9) pertaining to some particular layer are
absolutely equivalent and make an equal contribution to functional
$\mF$. Hence the spectrum of the eigenvalues in this mixed state
is degenerate, and degeneration multiplicity in each layer is
determined by value of $\|{\mathbf n}\|$ and space dimensionality
$s$
 $$
 g_s \left(\|{\mathbf n}\|\right)=\frac
 {\left(\|{\mathbf n}\| + s - 1\right)!}
 {\left(\|{\mathbf n}\|\right)!\left(s - 1\right)!}.
 $$
The total number of significant terms in (9) is given by relation
 $$
 \N(\cL)=\sum\limits_{\|{\mathbf n}\|=0}^{\cL-1}
 {g_s\left(\|{\mathbf n}\|\right)}=
 \frac{(\cL + s - 1)!}{(\cL - 1)!s!}
 = \frac{\cL}{s}g_s(\cL).
 $$
Finally, the sought state $\rho_\mathit{min}$, which realizes the
minimum of the generalized uncertainty relation, is described by
series (9) with coefficients
 $$
 a_{{\mathbf n},{\mathbf n}}=\left\{\begin{array}{l}
 \displaystyle
 \frac{1}{\N(\cL)}
 \left[1 + \left({(\cL - 1)s - \|{\mathbf n}\|(s + 1)}\right)
 \frac{\sqrt{(\N(\cL)-\Neff)(s + 2)}}
      {\sqrt{\Neff s(\cL + s)(\cL - 1)}}
 \right]\;
 ,0 \le \|{\mathbf n}\| < \cL\\
 0,\|{\mathbf n}\|\ge \cL\\
 \end{array} \right.,
 \eqno(13)
 $$
and the inequality analogous to (2) takes the form
 $$
 \Delta x\Delta q \ge
 \frac{(2\cL + s - 1)}{2(s + 1)} -
 \frac{\sqrt{(\N(\cL) - \Neff)(\cL + s)(\cL - 1)}}
      {(s + 1)\sqrt{\Neff s(s + 2)}}
 = B\left(\Neff,\cL(\Neff)\right).
 \eqno(14)
 $$
And it should be kept in mind that quantity $\cL$, a certain
positive integer, still remains a free parameter of the task and
should be chosen proceeding from the condition of minimality of
the right-hand side of inequality (14). By this means $\cL$ will
be a certain function of $\Neff$. The above requirements for the
expansion coefficients $a_{{\mathbf n},{\mathbf n}}$ (13) to be
real and positive impose restrictions on allowable values of this
parameter, namely, the quantity $\cL$ should satisfy the following
inequalities
 $$
 \Neff \le \frac{(\cL + s - 1)!}{(\cL - 1)!s!},
 \eqno(15)
 $$
 $$
 \Neff > \frac{(\cL + s - 1)!}{(\cL - 2)!(s + 1)!}
 \frac{s + 2}{s + 2(\cL - 1)}.
 \eqno(16)
 $$
There exist certain ranges of values of $\Neff$ and $s$ (in
particular, $s \gg 1$), in which inequalities (15), (16) determine
$\cL$ uniquely. But where it is not fulfilled and several integers
fall within the interval given by formulas (15) and (16), the
sought value of $\cL$ turns to be largest of them (i. e. integer
proximate to the upper boundary of (16)).

Inequality (14) is more hard than the previous inequality (2). On
the plane of parameters with coordinates $\Delta x\Delta q$ and
$\Neff$ it rigorously defines the region of physically realizable
states. It is therefore apparent and can easily be proved that one
should arrive at the same distributions of expansion coefficients
(13) and the same region (14) by solving the inverse problem of
seeking a mixed state with a largest possible value of $\Neff$ at
a given measure $\Dv$ of phase volume occupied by this state.

\section{Approximate form of uncertainty relation}

By virtue of the fact that with increasing $\Neff$ the quantity
$\cL$ goes through a set of discrete values, the boundary of
physical region $B(\Neff)$ in (14) is not a perfectly smooth curve
and is not described by analytical expression. Even though the
numerical calculation of the right-hand side of inequality (14)
presents no problem, it would be desirable to have its approximate
analytical form in order to analyse the obtained relation and
compare it with (2).

To this end we have investigated the behaviour of the function
$B(\Neff ,\tilde \cL)$ at a given value of $\Neff$, formally
regarding $\tilde \cL$ as an independent continuous variable.
Analytical and numerical calculations show that in the vicinity of
the true value of $\cL$ the dependence of $B$ on $\tilde \cL$ is
extremely weak. As well, in this region slightly different
neighbouring local minimum and maximum for $B(\tilde \cL)$ occur.
This gives grounds to use the continuous parameter $\tilde \cL$
instead of discrete number $\cL$ when approximately describing the
boundary of inequality (14). There are several ways of choosing
the specific condition determining the value of $\tilde \cL$.

The following one seems to be fairly simple and logically
substantiated. In the expression for expansion coefficients (13),
one can formally require ``continuous'' transition of its first
part to the second one\footnote{Naturally, in this case, the
variable $\|n\|$ in (13) should also be regarded as continuous.},
which, after a little algebraic manipulations, yields a
transcendental equation for determining $\tilde \cL$ in the form
 $$
 a_{\tilde \cL,\tilde \cL} = 0
 \Rightarrow \Neff = \frac{(s + 2)\,\Gamma(\tilde \cL + s + 1)}
                   {(s + 2\tilde \cL)(s + 1)!\,\Gamma(\tilde \cL)};
 \eqno(17)
 $$
(because of $\tilde \cL$ being continuous the factorials entering
into the formula for $\N(\cL)$ have been replaced here by Euler
gamma-functions). Accordingly, with such a way of defining $\cL$,
the approximate expression for inequality (14) can be written in
the parametric form
 $$
 \Delta x\Delta q \ge \frac{s + 2\tilde \cL(\Neff)}{2(s + 2)}.
 \eqno(18)
 $$
It should be noted that the approximate boundary of physical
domain, given by relations (17) and (18), exactly coincides with
the result of the procedure proposed in [9], proceeding from the
definition of $\cL$ as the upper bound of inequality (16).

Another way of finding a smooth form of the boundary, $B(\Neff)$,
can be chosen by substituting $\cL$ by the value of $\tilde \cL'$
corresponding to $\min_{\tilde \cL} B(\Neff, \tilde \cL)$. The
thus obtained curve has the advantage that it is wholly located in
the physical domain and is tangent to the strict boundary at
individual points. Unfortunately, at large $s$ this method does
not lead to analytical formulas.

Finally, according to the above mentioned property of equivalence
of the direct and the inverse problems, the domain of physically
realizable states can be represented as inequality to the largest
possible value of $\Neff$ at given $\Delta x\Delta q$. On the
basic of (17), (18) it gives the analytical expression
 $$
 \Neff \le \frac{1}{2\Delta x\Delta q}
  \frac{\Gamma \left( (s + 2)\Delta x\Delta q + s/2 + 1 \right)}
       {(s + 1)!\Gamma \left( (s + 2)\Delta x\Delta q - s/2 \right)}.
 \eqno(19)
 $$

The introduction of the concept of ``packing coefficient''
$C(s,\Neff)$ (whose explicit form is not described by simple
analytical formula) permits to rearrange relations (14) and (18)
into the form similar to (2). The point of such transformation is
that thus the main dependence of generalized uncertainty relation
on problem's parameters is emphasized whereas $C(s,\Neff)$ plays
role of correction factor with comparatively weak dependence on
$\Neff$.

The method of obtaining the approximate expression for the
physical domain boundary supposes itself that the approximation
(18) should tend to the exact formula (14) with increasing $\cL$
(and, thus $\Neff$ as well). Really, exactly in the limit of $\cL
\gg 1$ the substitution of a discrete value of $\cL$ by a
continuous value of $\tilde \cL$ slightly effects the weights of
individual states in (13). However, even at small $\Neff \gtrsim
1$ the approximate relations (18) and (19) turn to be in a very
good accord with the rigorous inequality (14). The reason for this
is the above mentioned local behavior of $B(\tilde \cL)$.

The said is illustrated by plots of exact and approximate
dependencies of $C(s,\Neff)$ given in the Fig. 1 for some values
of $s$. From this figure it is seen that insignificant
discrepancies between (14) and (18) only take place near $\Neff
\approx 1$ and they are the more pronounced, the greater the
problem dimension $s$. Accordingly, as $s$ is increased, thus
approximate relation approaches the rigorous one evenly closely at
larger values of $\Neff$.

Comparing inequalities (2) and (18), it can be easily shown that
(2) is an asymptotic form of (18) (and, consequently, (14)) at
$\Neff \rightarrow \infty$, but at $s>1$ a good approximation of
(2) to (14) is only attained at large $\Neff$. At the same time,
the refined form of the uncertainty relation (14) correctly
describes all range of values of $\Neff$ and, in particular, the
ultimate case of pure states
 $$
 C(s,\Neff = 1) = 1,\quad (\Delta x\Delta q)\Bigg|_{\Neff = 1}
 \ge \left(\frac{1}{2}\right)^s.
 $$

The coincidence of the asymptotic form of packing coefficient
$C(s,\Neff \gg 1)$ with formula (7) justifies the use in [5], when
proving (2), of assumption that the Wigner function is nonnegative
for the density matrix in the state with minimal uncertainty.

\section{Discussion}

At a qualitative level, the treatment of standard uncertainty
relation as a requirement for phase space quantization in wave
mechanics is rather frequently occurred in the literature on
physics (see, for example [11]). Therefore, it can serve, in some
way, as an argument in substantiating the postulate of statistical
mechanics on the number of cells in phase space of the system:
 $$
 \frac{{\mathcal V}_p (E)}{N} = (2\pi)^s,
 \eqno(20)
 $$
where ${\mathcal V}_p (E)$ is the phase volume of the system with
given energy $E$ and $N$ is the number of levels with energies not
exceeding $E$ (detailed definitions can be found in [4]). The
results of [1-3, 5, 9] and the present paper open up the
possibility of quantitative comparison of relations (2), (14) and
(20).

First of all their similarity lies in the fact that the quantities
of the same physical nature enter into the right- and the
left-hand sides of expressions (2) and (20) by identical manner.
But the differences between (2) and (20) are far more essential
and there is nothing strange about it, if we take into account the
way in which the concepts of the phase volume and the number of
states are introduced in both cases.

The postulate given by (20) specifies the relation between
energetic structure of the quantum system and the phase space
which can be associated with this system in the quasi-classical
approximation [4]. Such a relation is assumed to be universal and
independent of a particular Hamiltonian of the system. Formula
(20) describes system as a whole and has no connection to its
particular physical state.

On the contrary, inequalities (2), (14) refer just to the physical
state of quantum system. Like Eq. (20), they are explicitly
associated with neither the kind of Hamiltonian nor the structure
of energy levels. The only limitation is that since the basic
functions of expansion (9) form a complete set, the wave-packets,
satisfying the condition of minimum uncertainty for mixed states,
can be constructed only for Hamiltonians whose eigenfunctions also
form a complete orthonormal basis. This holds, in particular, for
the simplest system of noninteracting particles in a free space.
In general, however, the eigenstates of density operator are not
the states with constant energy and, thus, the state with minimum
uncertainty will not be stationary. The exception is the case of
multidimensional harmonic oscillator.

It is clear from the said that the quantities $\Dv$ and ${\mathcal
V}_p$, respectively, in Eq. (2), (14) and (20) not only do not
coincide numerically with one anther (even for oscillator), but
also have a different operational meaning. A similar statement in
general case is true for the quantities $\Neff$ and $N$ as well.
So there is nothing surprising in the fact that relation (2) has
the form of inequality, while (20) --- equality. Possibly, a
better correlation between (2) and (20) can be attained by using
other, alternative formulations of the uncertainty principle [9].

It would be worth to draw attention to one more difference between
(2) and (20). In the postulate given by (20), the phase space
volume of any quantum cell is constant and varies with change of
dimensionality in a strictly definite way --- as $({\mathrm
const})^s$. On the contrary, in the generalized uncertainty
relation (14), the minimal specific phase volume (i. e. the volume
per one effective pure state of the system) depends on the value
of $\Neff$. As seen from Fig. 1, the quantity $C(s,\Neff)$ is less
than unity and monotonically decreases with increasing $\Neff$. It
means that as the number of pure states involved in the
statistical mixture for $\rho ({\mathbf X},{\mathbf X'})$ raises,
there is a gain in the packing density of states.

In the paper [5] it was also noted the effect of increasing
packing with augment of the configurational space dimension $s$,
that is expressed by the inequality
 $$
 C(ks) \le C(s)^k ,\quad (k \ge 1).
 $$
From (19) one can easily see that an analogous relation for the
coefficients $C(s,\Neff)$ in the general case does not hold and
this property of packing is asymptotic, i. e. in each particular
case ($s$, $k$) it takes place beginning with sufficiently large
$\Neff$. The reason for this should be sought in the peculiarities
of behavior of degeneration multiplicity $g_s(\|{\mathbf n}\|)$ as
a function of $\|{\mathbf n}\|$ and $s$. It is natural to suppose
that the decrease in the coefficient $C(s)$ with increasing $s$ is
due to the rapid growth of the value of $g_s$, and for different
dimensions it would be proper to compare the packing at equal
values of the parameter $\cL$ rather than $\Neff$.

In this context, it is interesting to consider the case of the
density matrix (9) with equal weights of all pure states
$a_{{\mathbf n},{\mathbf n'}} = \delta _{{\bf n},{\bf n'}}/\N$
involved, that somewhat resembles the definition of $N$ in (20).
For this situation all methods of determining $\Neff$ gives $\Neff
= \N$ and, although the minimum of uncertainty relation (14) for
such a system is slightly exceeded, both the above properties of
the packing coefficient are fulfilled.

In conclusion, it is necessary to discuss also the question of
what physical meaning is attributed to the arguments (coordinates)
entering into the density matrix $\rho ({\mathbf X},{\mathbf
X'})$. These may be either the coordinates of one particle in the
real three-dimensional space or the coordinates of several
particles in the configurational space or both at once. The above
solution for the density matrix in the state with minimal
uncertainty is invariant under any permutation of its arguments
$(x_i  \Leftrightarrow x_j)$. For the real coordinates of one
particle it means that the corresponding wave-packet possesses
some rotational symmetry in three-dimensional space. But when
arguments being interchanged correspond to several particles it is
necessary to take into account the properties of their
permutational symmetry. For indistinguishable identical particles
the state with such a symmetry can only be realized in the case
when these particles follow the Bose statistics. On the contrary,
identical particles with Fermi statistics can not be described by
a density matrix of the form of (9) with the coefficients given by
(13). The state with minimal uncertainty for fermions should be
sought from the very beginning in the class of wavefunctions
antisymmetrized in permutations of arguments, which, naturally,
should lead to a result drastically different from (14).

\pagebreak

Figure caption

Fig. 1. The packing coefficient $C(s,\Neff)$ vs. the effective
number of pure states in statistical mixture (9) for dimensions $s
= 1,\,2,\,3$. Solid line --- strict inequality (14), dotted line
--- approximation (18), dashed line --- the asymptotic value
$C(s)$ for $\Neff \gg 1$ (7).


\end{document}